\documentclass[aps,preprint,groupedaddress]{revtex4-1}
\usepackage{graphicx}
\usepackage{graphics}
\usepackage{caption}
\usepackage{sidecap} 
\sidecaptionvpos{figure}{c}
\usepackage{wrapfig} 
\usepackage[toc,page]{appendix}
\usepackage{color}

\usepackage{amssymb,amsmath}
\begin{document}


\title{A mechanistic first--passage time framework for bacterial cell-division timing}



\author{Khem Raj Ghusinga}
\email[]{khem@udel.edu}
\affiliation{University of Delaware, Newark, Delaware, USA 19716.}
\author{Cesar A. Vargas-Garc\'ia}
\email[]{cavargar@udel.edu}
\affiliation{University of Delaware, Newark, Delaware, USA 19716.}

\author{Abhyudai Singh}
\email[]{absingh@udel.edu}
\homepage[]{http://udel.edu/~absingh/}
\affiliation{University of Delaware, Newark, Delaware, USA 19716.}


\date{\today}

\begin{abstract}How exponentially growing cells maintain size homeostasis is an important fundamental problem. Recent single-cell studies in prokaryotes have uncovered the adder principle, where cells on average, add a fixed size (volume) from birth to division. Interestingly, this added volume differs considerably among genetically-identical newborn cells with similar sizes suggesting a stochastic component in the timing of cell-division. To mechanistically explain the adder principle, we consider a time-keeper protein that begins to get stochastically expressed after cell birth at a rate proportional to the volume. Cell-division time is formulated as the first-passage time for protein copy numbers to hit a fixed threshold. Consistent with data, the model predicts that while the mean cell-division time decreases with increasing size of newborns, the noise in timing increases with size at birth. Intriguingly, our results show that the distribution of the volume added between successive cell-division events is independent of the newborn cell size. This was dramatically seen in experimental studies, where histograms of the added volume corresponding to different newborn sizes collapsed on top of each other. The model provides further insights consistent with experimental observations: the distributions of the added volume and the cell-division time when scaled by their respective means become invariant of the growth rate. Finally, we discuss various modifications to the proposed model that lead to deviations from the adder principle. In summary, our simple yet elegant model explains key experimental findings and suggests a mechanism for regulating both the mean and fluctuations in cell-division timing for size control.
\end{abstract}


\maketitle

\subsection*{Introduction}
One common theme underlying life across all organisms is recurring cycles of growth of a cell, and its subsequent division into two viable progenies. How an isogenic population of proliferating cells maintains a narrow distribution of cell size, a property known as cell size homeostasis, has been a topic of research for long time \cite{yfk75, fan77, wrp10,tes12a,mys12, am14, iwh14,rhk14, csk14,srb14,onl14, tbs15, jut15, dvx15, ro15}. Over the years, a number of theories have been proposed to explain cell size control from a phenomenological viewpoint. Recent single-cell experiments have shown that several prokaryotes such as \emph{Escherichia coli, Caulobacter crescentus, Bacillus subtilis}, \emph{Pseudomonas aeruginosa}, and \emph{Desulfovibrio vulgaris Hildenborough} \cite{csk14,tbs15, dvx15, fdm15} follow what is called an \emph{adder} principle. The adder model states that a cell, on average, adds a constant size between birth and division, irrespective of its size at birth \cite{vk97,iwh14,am14,csk14,srb14, jut15,tbs15,dvx15}. The adder mechanism has also been found to be present in the G1 phase of budding yeast \textit{Saccharomyces cerevisiae} \cite{srb14}, suggesting that it is possibly employed by a wide range of organisms. 

\par Despite a significant development in the understanding at phenomenological level, the molecular mechanisms underpinning the cell size control are not well understood \cite{DoB03,ro15}. A prevalent class of models posits that a protein acts as a time-keeper between subsequent occurrences of an important event in the cell cycle \cite{tcd74,fan75, onl14, dvx15, bzd15, do68, sm73, am14, ha15}. This protein is synthesized at a rate proportional to instantaneous volume (size) and the event of interest, which could either be the division itself \cite{tcd74,fan75, onl14, dvx15, bzd15} or commitment to division after a constant time \cite{do68, sm73, am14, ha15}, takes place when the protein reaches a certain threshold. It has been previously shown that this simple biophysical mechanism can lead to the adder principle of cell size control in the mean sense \cite{am14,ha15,bzd15}. Interestingly, recent experimental data on bacteria has quite fascinating stochastic component as well. For instance, not only the mean but the distribution of the volume added between two division events itself is independent of the volume at birth for a given growth condition. Also, in different growth conditions, the distributions of the added volume collapse if scaled by their respective means \cite{tbs15}. It remains to be seen whether these stochastic traits can be produced by the aforementioned molecular mechanism. 

A plausible source of the stochasticity is the noise in gene expression wherein randomness in transcription and translation results in significant cell to cell variation in protein levels \cite{bkc03, rao05, rav08,keb05, sis13}. Here, we consider a time-keeper protein between two consecutive division events and show that its probabilistic expression can indeed manifest as the stochastic component in cell size.  An overview of the mechanism and it leading to added volume distribution independent of the initial cell volume are depicted in Fig.~\ref{fig:mechanism}. Considering a cell of given volume at birth, we assume that its volume grows exponentially until the time at which the time-keeper protein triggers the division.  The production of the protein is started right after the birth of a cell. The production rate of the protein scales with the volume, i.e., increases exponentially (Fig.~\ref{fig:mechanism}(a)). Cell division time is modeled as the first-passage time for protein copy numbers to attain a certain threshold (Fig.~\ref{fig:mechanism}(b)). The distribution of the first-passage time is then used to compute the distribution of the volume added between birth to division (Fig.~\ref{fig:mechanism}(c),(d)). Our results further show that distributions of the added volume, and cell division time have scale-invariant forms: they collapse upon rescaling with their respective means in different growth conditions. Lastly, we discuss implications of these findings in identification of the time-keeper protein. We also deliberate upon various modifications to the proposed model that result in deviations from the adder principle. 

\begin{figure}[b]
\centering
\includegraphics[width=0.65\linewidth]{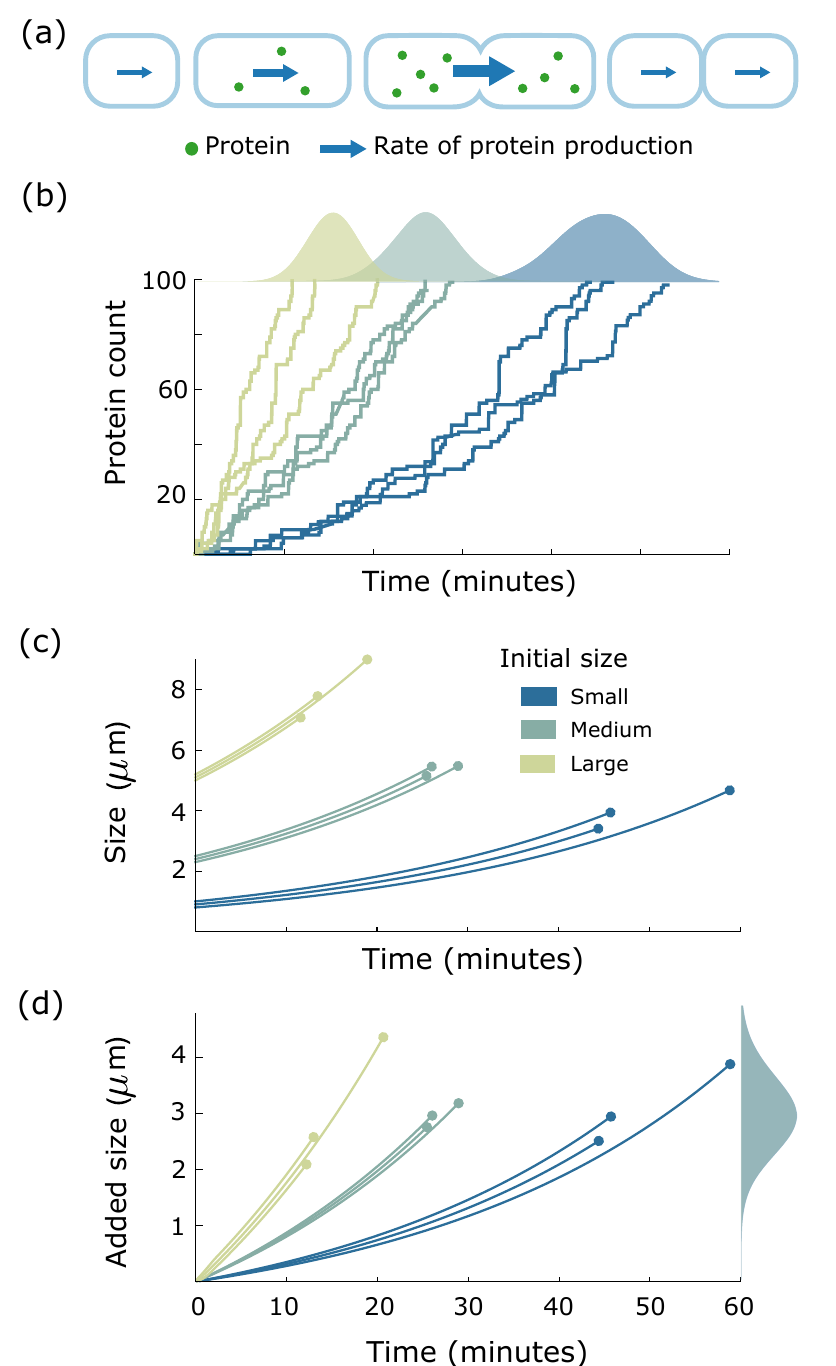}
\caption{\emph{A molecular mechanism that explains adder principle}. \textbf{(a)} A rod-shaped growing cell starts synthesizing a protein right after its birth. Th production rate of the protein scales with the size (volume) of the cell. When the protein's copy number attains a certain level, the cell divides and the protein is degraded.}
\label{fig:mechanism}
\end{figure}

\begin{figure}[!ht]
\ContinuedFloat
\caption{\textbf{(b)} The stochastic evolution of the protein is shown for cells of three different sizes at birth. Each cell divides when the protein's level achieves a specific threshold.  The distribution of the first-passage time  (generated via 1000 realizations of the process) for each newborn cell volume is shown above the three corresponding trajectories. The first-passage time distribution is dependent upon the newborn cell size; on average the protein in a smaller cell takes more time to reach the threshold as compared to the protein in  a larger cell. 
\textbf{(c)} The time evolution of size is shown for cells of different initial volume. The size is assumed to grow exponentially until the protein in the top figure reaches a critical threshold. The division is assumed to take place then. \textbf{(d)} The size added to the initial size when the division event takes place is shown for cells of different volume. The distribution of the added size is independent of the initial size of the cell, thus reproducing an adder model.}
\label{fig:mechanism}
\end{figure}

\subsection*{Model description}
Considering a cell with volume at birth $V_0$, its volume at a time $t$ after birth is given by
\begin{equation}
V(t)=V_0 \exp(\alpha t),
\end{equation} 
where $\alpha$ represents cell growth rate. As shown in Fig.~\ref{fig:mechanism}(a), production of a protein is started right after the birth of a cell. A stochastic gene expression model describing the time evolution of this protein is described below.

Let $x(t)$ denote the protein count at time $t$. Assuming constitutive transcription and  mRNA half life is smaller than the cell cycle time, we make the translation burst approximation where each mRNA molecule degrades instantaneously after producing a burst of protein molecules \cite{Pau05,FCX06,ShS08,Ber78,EJF11,Rig79}. Protein  synthesis is given by the following model:
\begin{equation}
Gene\xrightarrow{r(t)} Gene + B_i \times Protein.
\label{eqn:GeneExpModel}
\end{equation}
The variable $B_i$ denotes the size of $i^{th}$ protein burst which, for each $i \in \{1, 2, 3, \cdots \}$, is independently drawn from a positive-valued distribution. It essentially represents the number of protein molecules synthesized in a single mRNA lifetime and typically follows a geometric distribution \cite{Ber78, Rig79, FCX06, SRC10, SiB12, SRD12}. Furthermore, the production rate $r(t)$ of the protein or alternatively the burst arrival rate (transcription rate) is assumed to be volume dependent. More specifically, we consider $r(t)$ as
\begin{equation}
r(t)=k_{m} V(t)=k_{m} V_0 e^{\alpha t},
\label{eqn:TranscriptionRate}
\end{equation} 
where $k_{m}$ is a proportionality constant. In this formulation, the production rate scaling with the cell volume is an essential component of maintaining concentration homeostasis. Indeed such a dependency of transcription rate on cell volume has been observed in mammalian cells \cite{pnb15}. 

Note that the protein count 
\begin{equation}
x(t)=\sum_{i=i}^n B_i, \quad \left<B_i\right>:=b,
\end{equation}
 is a sum of independent and identically distributed random variables $B_i$'s, where $n$ is the number of bursts arrived (transcription events) in time interval $[0,t]$. The key assumption  is that the cell divides when the protein level $x(t)$ crosses a threshold $X$. We would like to mention that though the model described here is for gene expression, it can be used to model any process involving accumulation of molecules wherein the production rate scales with the volume. A general distribution for $B_i$ allows a wider range of processes to be covered. The scope can be further widened by considering the parameters $k_m$, $b$, and $X$ to be functions of the growth rate $\alpha$. In the next section, we quantify the cell division time by modeling it as the first time taken by $x(t)$ to cross $X$. 

\subsection*{Cell division time as a first-passage time problem}
The first-passage time ($FPT$) for the stochastic process $x(t)$ to cross a threshold $X$ is mathematically defined as:
\begin{equation}
FPT:=\inf\left\{ t:x(t)\geq X | x(0)=0\right\}\text{.}
\label{eqn:FPT}
\end{equation}
For the stochastic gene expression model described in the previous section, time evolution of $x(t)$ and corresponding first-passage times for cells with different initial volumes is depicted in Fig.~\ref{fig:mechanism}(b). To characterize the distribution of $FPT$, we need to quantify two quantities: arrival time $T_n$ of a $n^{th}$ burst, and minimum number of burst (transcription) events $N$ required for $x(t)$ to cross $X$. The conditional probability density function of $FPT$ for a given cell volume at birth $V_0$ is then given as
\begin{align}
f_{FPT|V_0}(t)= \sum_{n=1}^{\infty} f_{T_n}\left(t\right) f_N(n),
\label{eqn:FPTpdfV0}
\end{align}
where $f_{T_n}(t)$ represents the probability density function of $T_n$, and $f_N(n)$ represents the probability mass function of $N$. 

Since $x(t)$ can only increase in time, the distribution of $N$ can be determined by the distribution of the burst size $B$ using the relation:
\begin{equation}
Prob\left(N \leq n\right)=Prob\left(\sum_{i=1}^n B_i \geq X\right).
\label{eqn:relnNB}
\end{equation}
As an specific yet physiologically relevant example, when the burst size distribution is considered to be geometric, $f_N(n)$ is given by
\begin{equation}
f_{N}(n)={n+X-2 \choose n-1}\left(\frac{1}{b+1}\right)^{n-1}\left(\frac{b}{b+1} \right)^X,
\label{eqn:distN}
\end{equation}
where $b$ represents the mean (or expected value) of burst size $B_i$ \cite{GhS14,SiD14}. Furthermore, the probability density function of $n^{th}$ arrival event  $f_{T_n}$ is governed by the underlying burst arrival process. In our case, the transcription (burst arrival) rate is time dependent. Therefore, the burst arrival process is an inhomogeneous Poisson process for which the probability density function of the arrival times are available in literature (see \cite{Bax82, PSZ00}). In particular, we have
\begin{equation}
f_{T_n}(t)=\frac{\left(R(t)\right)^{n-1}}{(m-1)!} r(t) \exp(-R(t)).
\label{eqn:Tnpdf}
\end{equation}
The transcription or burst arrival rate $r(t)$ is referred to as the intensity function of the corresponding inhomogeneous Poisson process. Also, $R(t)$ is the mean value function of the inhomogeneous poisson process
\begin{equation}
R(t):=\int_{0}^{t}r(s)ds=\frac{k_{m} V_0}{\alpha}\left(e^{\alpha t}-1\right).
\end{equation}

\begin{figure}[hbtp!]
\centering
\includegraphics[width=\linewidth]{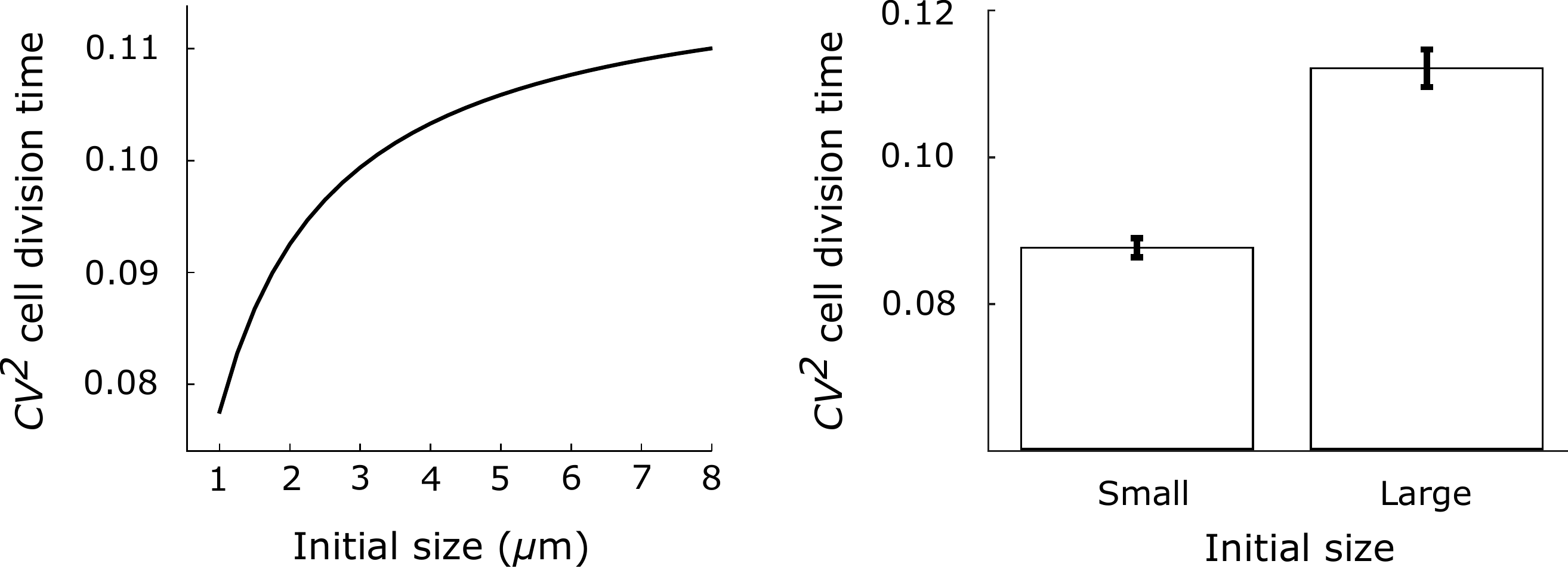}
\caption{\emph{The noise in the cell division time increases with initial volume}. \textbf{Left}: The noise in division time ($CV^2$ of $FPT$) increases as the cell volume at birth $V_0$ is increased. Using the expression of the first-passage time in \eqref{eqn:FPTpdf}, the noise ($CV^2$) is numerically computed for different values of the newborn cell volume. The parameters used  for the model are $k_{m}=1$ (1/min), $X=100$ (molecules), $\alpha=0.03$ (1/min), and geometric distribution of the burst size $B$ with mean $b=3$ (molecules). \textbf{Right}:  The prediction from the model is validated using the experimental data from \cite{tbs15}. Using bootstrapping, a statistically significant (p-value=$0.00023$) increase in the noise is observed from smaller initial volume to a larger initial volume (more details in the Supplementary Information, section S5).}
\label{fig:noiseFPT}
\end{figure}

Using \eqref{eqn:Tnpdf} in \eqref{eqn:FPTpdfV0}, the expression for the probability density function of $FPT$ for a cell of given volume at birth becomes
\begin{equation}
f_{FPT|V_0}(t)=\sum_{n=1}^{\infty} \frac{\left(R(t)\right)^{n-1}}{(n-1)!} r(t) \exp(-R(t)) f_N(n).
\label{eqn:FPTpdf}
\end{equation}
The conditional $FPT$ distribution in \eqref{eqn:FPTpdf} qualitatively emulates the experimental observations in  \cite{tbs15} that the mean cell division time decreases as the cell size at birth is increased (refer to SI, section S1). This is intuitively expected as a cell with large size a birth will have a higher transcription rate as compared to a cell with small size. Hence, on average, the time taken by the protein to reach the prescribed threshold is smaller in the larger cell.  The model also predicts that the noise (quantified using coefficient of variation squared, $CV^2$) in the cell division time increases as new born cell volume is increased (see Fig. \ref{fig:noiseFPT} (left)). This prediction is consistent with data from \cite{tbs15}, as shown on the right part of Fig. \ref{fig:noiseFPT} . The noise behavior can be explained by observing that on average a cell with smaller volume at birth takes more time for division. Therefore the fluctuations are time averaged, leading to a smaller noise in division time as compared to that of a cell with larger volume at birth.

\subsection*{Distribution of the volume added between divisions}
In the previous section, we determined the distribution of the division time for a cell of given initial volume $V_0$. Coupling this with the fact that volume of the cell grows exponentially in time, we can find the distribution of the volume added to $V_0$. More precisely, the volume added from birth to division (denoted by $\Delta V$) is given as 
\begin{equation}
\Delta V = V_0 \left( e^{\alpha FPT}-1\right)\text{.}
\end{equation}
The distribution $FPT$ given in \eqref{eqn:FPTpdf} can be used to find the distribution of $\Delta V$ (see SI, section S2). Let $f_{\Delta V}(v)$ denote the probability density function of $\Delta V$, then we have
\begin{align}
f_{\Delta V}(v)=\sum_{n=1}^{\infty} \frac{\left(\frac{k_{m} v}{\alpha}\right)^{n-1}}{(n-1)!} \frac{k_{m}}{\alpha} \exp\left(-\frac{k_{m} v}{\alpha}\right) f_N(n)\text{.}
\label{eqn:dVpdf}
\end{align}
One striking observation is that $f_{\Delta V}(v)$ is independent of the initial volume $V_0$ (see Fig~\ref{fig:mechanism} for a depiction of this). This is in agreement with the experimental observations that the distribution of the added volume does not depend on the size of the cell at birth \cite{iwh14,tbs15}.  Further, we can also use the expression in \eqref{eqn:dVpdf} to find moments of $\Delta V$. We will first discuss the expression of mean $\Delta V$, particularly emphasizing its dependence on the growth rate. The higher order moments are taken up next. 

\subsubsection*{Mean of added volume}
The distribution of $\Delta V$ given in \eqref{eqn:dVpdf} is an Erlang distribution conditioned on $N$, the minimum number of burst events required for the protein to cross the threshold. The formula for $\left<\Delta V\right>$ can be written as
\begin{equation}
\left<\Delta V\right>=\frac{\alpha}{k_{m}} \left<N\right>=\frac{\alpha}{k_{m}} \left(\frac{X}{b}+1\right),
\label{eqn:deltaVmean}
\end{equation}
where we have assumed that protein is produced in geometric bursts with mean burst size $b$ (see SI, sections S1, S2). It can be seen that the if we consider the parameters $k_m$, $b$, and $X$ to be independent of the growth rate $\alpha$, average volume added is a linearly increasing function of $\alpha$ . This is consistent with the analysis and experimental data on \textit{Pseudomonas aeruginosa} \cite{dvx15}. In contrast, other studies have mentioned that there is an exponential relationship between $\left<\Delta V\right>$ and $\alpha$  instead \cite{csk14,srb14,tbs15}. This nebulous aspect of the data has been attributed to narrow range of achievable growth rates which makes it difficult to discern a linear dependency from an exponential one \cite{dvx15}. While suitable forms of $k_{m}$, $b$, and $X$ as functions of $\alpha$ can generate a desired growth rate dependency of the added volume, we discuss a modification in the model to see an exponential relationship between them for the case when $k_m$, $b$, and $X$ are constants. 

Let us consider that instead of accounting for the time between birth to division, the protein accounts for time between other two events in the cell cycle. More specifically, we consider initiation of DNA replication takes place when sufficient time-keeper protein has been accumulated per origin of replication \cite{do68,am14,ha15}. The corresponding division event is assumed to occur with a constant delay of $T$ after an initiation. Here the delay $T$ is what is called $C+D$ period whereby $C$ represents the time to replicate the DNA and $D$ denotes the time between DNA replication and division \cite{CoH68,Coo12}. As shown in \cite{ha15} (section 3.2), the volume added between two consecutive initiation events for each origin of replication is same as $\Delta V$ in our previous model. Further, the average volume added between divisions associated associated with these initiations (denoted by $\Delta V^{*}$) is related with $\Delta V$ as
\begin{equation}
\left<\Delta V^{*}\right> \approx \left<\Delta V \right> e^{\alpha T} = \frac{\alpha}{k_{m}} \left(\frac{X}{b}+1\right) e^{\alpha T}.
\label{eqn:deltaVdiv}
\end{equation}

When $k_m$, $b$, and $X$ are constants, the expression in \eqref{eqn:deltaVdiv} suggests two different regimes of how $\Delta V_{div}$ depends upon $\alpha$. For small values of $\alpha$, we have $\alpha \exp(\alpha T) \approx \alpha$. Therefore the mean added volume increases linearly with the growth rate. For large enough values of $\alpha$, the exponential term starts dominating which leads to exponential dependence of the average volume added with change in the growth rate. This also means that if the growth rate $\alpha$ is small, it is not possible to distinguish whether the underlying mechanism accounts for volume added between two division events or two initiation events as the data will show linear dependence of the average added volume with changes in $\alpha$ \cite{dvx15}. Notice that a pure exponential relationship between $\Delta V^{*}$ and $\alpha$ can also be obtained by having $k_{m}$ a linearly increasing function of $\alpha$. For this particular case, the volume accounted for by each origin of replication $\Delta V$ will become invariant of the growth rate as suggested in \cite{ha15,tah15}. 
 
 To sum up, the case when the protein accounts for volume added between two divisions gives a linear dependency of the mean added volume on growth rate. However, an exponential dependence between these quantities can be achieved by having the protein account for volume added between two initiation events. We now go back to discussing the higher order moments of the division to division model.

\subsubsection*{Higher order moments of added volume and scale invariance of distributions}
We can use the distribution of $\Delta V$ to get its higher order statistics (see SI, section S2). In particular when the protein production is considered in geometric bursts, the coefficient of variation squared ($CV^2_{\Delta V}$) and skewness ($skew_{\Delta V}$) are given by 
\begin{align}
CV^2_{\Delta V}= \frac{b^2+2 b X+X}{(b+X)^2}, \quad skew_{\Delta V}= \frac{2 \left(b^3+3 b^2 X+3 b X+X\right)}{\left(b^2+2 b X+X\right)^{3/2}}.
\label{eqn:CVSkewDeltaV}
\end{align}  
These formulas show that both $CV^2$ and skewness do not depend on the growth rate $\alpha$. It turns out an even more general property is true: an appropriately scaled $j^{th}$ order moment of  $\Delta V$, i.e., $\left<\Delta V^j\right>/\left<\Delta V\right>^j$ is independent $\alpha$. This arises from the fact that the distribution of $\Delta V$ can be written in the following form
\begin{equation}
f_{\Delta V}(v)=\frac{1}{\left<\Delta V\right>} g \left( \frac{v}{\left<\Delta V\right>}\right),
\end{equation}
regardless of the distribution of the burst size $B$. An important implication of above form of distribution of $\Delta V$ is the scale invariance property: the shape of the distribution across different growth rates is essentially same, and a single parameter $\left<\Delta V\right>$ is sufficient to characterize the distribution of $\Delta V$ \cite{gac13}. Recent experimental data have also exhibited the scale invariance property \cite{tbs15,koj14,bzd15}.

Interestingly, the above invariance property is not limited to the distribution of the added volume $\Delta V$. Ignoring the partitioning errors in the volume, it can be seen that in steady-state the cell-size distribution at birth is approximately same as the distribution of $\Delta V$ \cite{tbs15}. Also, the size at division is $2 \Delta V$. Thus, the scale invariance of $\Delta V$ immediately implies scale invariance of the distributions of cell sizes at birth and division \cite{tbs15}. Moreover, the distribution of the division time can also be determined by unconditioning \eqref{eqn:FPTpdf} with respect to the distribution of the initial volume $V_0$. As shown in the SI (section S3), the distribution of the division time also has the scale invariance property which is in agreement with the results in \cite{ics14,iwh14}.

\subsection*{Discussion}
In this work, we studied a molecular mechanism that can lead to adder principle of cell size control. This mechanism imagines a protein sensing the volume added between birth to division  \cite{tcd74,fan75, onl14, dvx15, bzd15}  or two other events in the cell cycle \cite{do68, sm73, am14, ha15}. Our work shows that this mechanism can exhibit the stochastic traits observed in the data \cite{tbs15}. In particular, it is shown that the distribution of volume added between birth to division is independent of the initial cell volume  \cite{tbs15}. Further, the distributions of key quantities such as the added volume, division time, volume at birth, volume at division, etc. show the scale invariance property  \cite{tbs15}. Our study also revealed that the noise in division time increases with increase in cell size at birth, which was validated from the available data from  \cite{tbs15}. Here, we discuss the implications of these results.

\subsubsection*{Potential candidates for the time-keeper protein}

Among many proteins involved in the process, prominent candidates for the time-keeper are FtsZ and DnaA. More specifically,  if the constant volume addition is considered between division to division, FtsZ  is a potential candidate for the protein \cite{bil90,lut07,eao10, deb10, ro15}. It plays an important role in determination of the timing of cell division which is triggered upon assembly of FtsZ into a ring structure \cite{ade09}. It has been proposed that the accumulation of FtsZ up to a critical level is required for cell division \cite{bil90, chl12}. Likewise, the protein DnaA is known to regulate the timing of initiation of replication, thus presenting a strong candidature for the protein if the constant volume is added between two initiation events \cite{alh87, lsh89,ro15}. In this case, initiation is thought to occur when a critical number of DnaA-ATP molecules are available \cite{DoB03}. Upon initiation, these DnaA-ATP molecules get deactivated by converting to DnaA-ADP \cite{sbk87, DoB03}.  While it is not clear yet whether the production of DnaA or its conversion to DnaA-ATP is a rate limiting step in the initiation process, the model presented here can account for both cases as long as the conversion to DnaA-ATP happens at a volume dependent rate. 

We can employ the closed-form expressions for the moments of $\Delta V$ developed in this work to investigate roles of these candidate proteins. Considering geometrically distributed burst of proteins, the expressions of mean and coefficient of variation squared ($CV^2$) are given by \eqref{eqn:deltaVmean} and \eqref{eqn:CVSkewDeltaV} respectively. Thus, increasing the threshold $X$ or decreasing the mean burst size of the time-keeper protein $b$ should result in decrease in the $CV^2$ of the added volume. Experimentally, the mean burst size can be altered by changing the translation rate of the proteins using techniques such as mutations in the Shine-Dalgarno sequence. Changing the threshold can be achieved by changing the protein sequence which affects its function and thus leads to a different number of protein molecules being required for division. It is important to point out cell-size control can possibly have mechanisms in place to overrule such tweaking. One possible way to overcome this could be to appropriately change the transcription rate scaling factor $k_{m}$ by promoter mutations, along with changes in $b$ or $X$ such that the added volume is same in the mean-sense. 

We also acknowledge that a complex process like cell division may have a lot more going on than a simple protein carrying out the size and time control. For instance, there is some evidence of DnaA not being solely responsible for the timing of initiation \cite{fft15}, a cell compensating for a larger or smaller initiation time by adjusting the genome replication time $C$ \cite{bsk96,hkc12,chl12}, etc. Along the same lines, FtsZ ring formation is inhibited upon DNA damage which suggests that a viable copy of DNA is required for division to proceed \cite{mcl98, mhl11}. It appears that several key proteins follow the dynamics of the hypothetical protein we considered and at important stages, check points are established for proper coordination. Nonetheless, our model can shed light into how perturbations in expression of these proteins can lead to experimentally observable changes. This provides exciting avenues for investigating different candidate proteins by examining the effect of alterations in gene expression parameters.

\subsubsection*{Other sources of noise}
The source of noise accounted for in this work is the intrinstic noise arising because of random birth events of mRNA/protein molecules, and death of mRNA molecules. In principle, there are other sources of noise such as cell-to-cell variation in cell specific factors such as enzyme levels which could affect the expression of the time-keeper protein and, in turn, influence the distributions of division time, cell size, etc. It is also possible that noise arising out of other factors dominates the noise from stochastic expression. 

One important parameter in our model is the event threshold $X$ which we have assumed to be fixed. It is possible that instead of a strict requirement of exactly $X$ molecules, the division event has an increased propensity as the protein count $x(t)$ gets closer to $X$. Our analysis in SI, section 4 shows that in order to get $\left<\Delta V\right>$ independent of the cell volume at birth $V_0$, the propensity of division requires a strict attainment of $X$ molecules. Thus, the alternate mechanism can be ruled out.

Recall our discussion in prevision section that $CV^2_{\Delta V}$ decreases as the threshold $X$ is increased. Thus for a very large threshold, the contribution from expression of the protein is negligible. To get a $CV$ of $\sim 20\%$ without other factors being counted in, we need a threshold of about  $20$ molecules. Interestingly, the number of DnaA-ATP molecules required for initiation are around $20$ \cite{DoB03}. The threshold for FtsZ, however, is thought to be somewhere between $4000$ molecules \cite{rvm03} to $15000$ molecules \cite{lse98}. Therefore the stochastic expression of protein suffices to account for noise in $\Delta V$ if the initiation to initiation mechanism via DnaA is the key regulator of cell cycle. However, it becomes inevitable that other sources of noise are also considered in the model if the regulation is from division to division via FtsZ. One possibility is to consider the cell-to-cell variations in the growth rate. Likewise, in the case when the protein accounts for the volume added between two initiation events, partitioning errors can be introduced upon division of a cell. Accounting for these factors would provide a better insight into the process. 

\subsubsection*{Deviations from the adder principle}

Recently it has been proposed that cells employ a generalized version of the adder principle wherein the volume added between divisions depends upon the cell volume at birth \cite{jut15,tpp15}. An important implication of this is on the time a cell will take to converge to steady-state value. For example if the added volume decreases with $V_0$ then a large cell will converge faster than it would have in a perfect adder strategy.

There could be several ways to get deviations from the adder principle. For instance, if we consider that the time-keeper protein does not degrade fully upon division and the remaining proteins are divided in the daughter cells in proportion to their respective volumes at birth, the added volume decreases as volume of daughter cell is increased. This is because if there are already time-keeper proteins present in the cell at the time of its birth, the threshold will be achieved earlier than the case when there were no proteins at birth. As a result, the added volume will be smaller as compared to an adder principle. Another possible way of getting such deviation could be if the mean burst size is an increasing function of the cell volume. It could also be explained by similar reasoning that it leads to a smaller time to reach the copy number threshold of the protein. In contrast, to get a higher added volume for a increase in $V_0$, we can curb the scaling of protein accumulation with the cell volume. One example is to assume a transcription rate of the form $r(t)=k_{m} \frac{V(t)}{V(t)+\overline{V}}$ where the transcription rate saturates with increase in volume. Alternatively, this effect could also be achieved by considering that instead of cell division occurring upon achieving a constant volume addition, its propensity increases as the added volume increases. 
\subsection*{Summary}
This paper shows that the stochastic accumulation of a time-keeper protein can lead to adder principle of cell size control. We derived analytical formulas for the division time and volume added between birth to division. These expressions were used to show that the volume added is independent of the cell size at birth, consistent with experimental data. Furthermore, the distributions of added volume and division time also show scale invariance property wherein the distribution can be uniquely determined by its mean in a given growth condition. We also discussed the implications of these results in identifying a possible molecular mechanism underlying the cell-size control. Finally, we discussed how the proposed mechanism can be modified to get more general behaviors. Future work will involve accounting for other sources of noise such as growth rate fluctuations, partitioning errors, etc.
 
\newpage
\onecolumngrid
\section*{Supplementary Information}
\renewcommand{\thesection}{S\arabic{section}}
\renewcommand{\thesubsection}{\thesection-\alph{subsection}}
\numberwithin{equation}{section}
\numberwithin{figure}{section}

 \section{Remarks on the conditional distribution of $FPT$ given cell volume at birth}
 As discussed in \eqref{eqn:FPTpdf} in the main text, the probability density function of the first-passage time given cell volume at birth $V_0$ is given by
\begin{equation}
f_{FPT|V_0}(t)=\sum_{n=1}^{\infty} \frac{\left(R(t)\right)^{n-1}}{(n-1)!} r(t) \exp(-R(t)) f_N(n).
\label{eqn:FPTdistSI}
\end{equation}
Here, $f_N(n)$ represents the probability mass function of $N$ (the minimum number of transcription events required for protein level to cross a threshold $X$). The relation in \eqref{eqn:relnNB} can be used to quantify the distribution $N$ from the distribution of protein burst size $B_i$. We present the form of distribution of $N$ for two relevant cases here: when protein burst size is one with probability one, and when the protein burst size is geometric \cite{Ber78, Rig79, FCX06, SRC10, SiB12, SRD12}.

When the burst size $B_i$ is one with probability one, \textit{exactly} $X$ events are required for the protein level $x(t)$ to reach $X$ for the first time. That is, we have
\begin{equation}
f_N(n)=\delta(n-X),
\end{equation}
where $\delta(n-X)$ is the Kronecker's delta which is one when $n=X$ and zero otherwise.  

For the case where the burst size $B_i$ follows a geometric distribution  \cite{Ber78, Rig79, FCX06, SRC10, SiB12, SRD12}, the calculation of the minimum number of transcription events $N$ for this distribution has been previously done in our works \cite{GhS14, SiD14}. The probability mass function of $N$ is given by
\begin{equation}
f_{N}(n)={n+X-2 \choose n-1}\left(\frac{1}{b+1}\right)^{n-1}\left(\frac{b}{b+1}\right )^X.
\end{equation}
Here $b$ represents the mean protein burst size. Further, the first three statistical moments of $N$ given by the above probability mass function are
\begin{align}
&\left<N\right>=\frac{X}{b}+1, \\
&\left<N^2\right>=\frac{b^2+3bX+X+X^2}{b^2}, \\
&\left<N^3\right>=\frac{b^3+7 b^2 X+6 b X (X+1)+X \left(X^2+3 X+2\right)}{b^3}.
\end{align}
\subsection*{Mean FPT given newborn size}
\begin{figure}[h]
\centering
\includegraphics[width=0.45\textwidth]{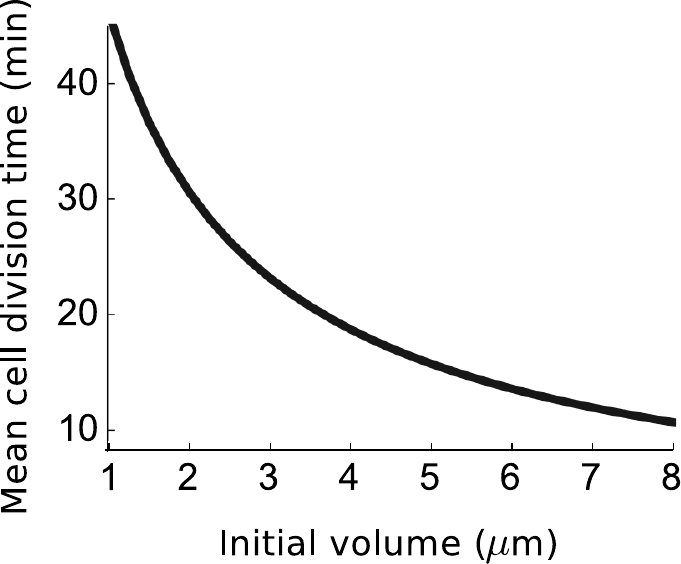}
\caption{The division time decreases as the cell size at birth increases. The division time (mean $FPT$) given a newborn volume  is computed numerically for each value of initial cell volume $V_0$. The protein production is assumed in geometric bursts and the parameters assumed are: $k_{m}$=1 per minute, X=100 molecules, $b$=3 molecules, $\alpha$=0.03 per minute.}
\label{fig:fptvsv0}
\end{figure}

The expression of $FPT$ probability density function in \eqref{eqn:FPTdistSI} can be used to determine the mean numerically. As we mentioned in the main text that, consistent with experiments, the mean $FPT$ decreases with increase in the volume of the cell at birth  ($V_0$). That is, a large cell on average divides earlier than a small cell. This is depicted in Fig. \ref{fig:fptvsv0}. 
\section{Distribution of $\Delta V$}
Let us assume that the initial volume of a (newborn) cell is $V_0$. Our primary assumption is that the cell divides at the first-passage time we computed in the previous section. Representing the volume added to the cell's volume at birth until the event takes place by $\Delta V$, we have 
\begin{equation}
\Delta V = V_0 \left( e^{\alpha FPT}-1\right).
\end{equation}
We already have the distribution of $FPT$ in \eqref{eqn:FPTdistSI}. We use it to determine the distribution of $\Delta V$ as follows.
\begin{align}
Prob \{ \Delta V \leq  v \} &= Prob \left\{ V_0 \left( e^{\alpha FPT}-1\right) \leq v\right\} \\
                                                 & = Prob \left\{ FPT \leq \frac{1}{\alpha}\ln \left(\frac{v}{V_0}+1\right)\right\} \\
                                                 &= \int_{0}^{\frac{1}{\alpha}\ln \left(\frac{ v}{V_0}+1\right)} f_{FPT|V_0}(t) dt.
\end{align}
Therefore the probability density function of $\Delta V$ is given by
\begin{align}
f_{\Delta V}(v)=& \frac{d}{d v} \left(Prob \{ \Delta V \leq  v \}\right) \\
                      =& \frac{d}{d v}  \int_{0}^{\frac{1}{\alpha}\ln \left(\frac{ v}{V_0}+1\right)} f_{FPT}(t) dt \\
                      =& f_{FPT|V_0}\left(\frac{1}{\alpha}\ln \left(\frac{ v}{V_0}+1\right)\right) \frac{d}{dv}\left(\frac{1}{\alpha}\ln \left(\frac{ v}{V_0}+1\right)\right).
\label{eqn:deltaVdistSI}
\end{align}
Note that $\displaystyle \frac{d}{dv}\left(\frac{1}{\alpha}\ln \left(\frac{ v}{V_0}+1\right)\right) = \frac{1}{\alpha \left(V_0+v\right)}$. Also
\begin{equation*}
 R\left(\frac{1}{\alpha}\ln \left(\frac{ v}{V_0}+1\right)\right)=\frac{k_{m} v}{\alpha}, \quad r\left(\frac{1}{\alpha}\ln \left(\frac{ v}{V_0}+1\right)\right)=k_{m}
\left(v+V_0\right).
\end{equation*}
Hence we can write the probability density of $\Delta V$ as 
\begin{align}
f_{\Delta V}(v)=&\frac{1}{\alpha \left(V_0+v\right)} \sum_{n=1}^{\infty} \frac{\left(\frac{k_{m} v}{\alpha}\right)^{n-1}}{(n-1)!} k_{m}
\left(v+V_0\right)\exp\left(-\frac{k_{m} v}{\alpha}\right) f_N(n) \\
=&  \sum_{n=1}^{\infty} \frac{\left(\frac{k_{m} v}{\alpha}\right)^{n-1}}{(n-1)!} \frac{k_{m}}{\alpha} \exp\left(-\frac{k_{m} v}{\alpha}\right) f_N(n).
\label{eqn:distDeltaV}
\end{align}
Notice that this distribution is an Erlang distribution conditioned to the random variable $N$.

\subsection*{Moments of $\Delta V$}
\paragraph{Mean $\Delta V$}
Since the distribution of $\Delta V$ is conditional Erlang, we have the following expression for mean $\Delta V$.
\begin{align}
\left<\Delta V\right> = \sum_{n=1}^{\infty} \frac{n}{k_{m}/\alpha}f_N(n)=\frac{\alpha}{k_{m}}\left<N\right>.
\label{eqn:meanDeltaV}
\end{align}
\paragraph{Second order moment} The second order moment of $\Delta V$ is given by
\begin{align}
\left<\Delta V^2\right> = \sum_{n=1}^{\infty} \frac{n^2+n}{\left(k_{m}/\alpha\right)^2}f_N(n)=\frac{\alpha^2}{k_{m}^2}\left(\left<N^2\right>+\left<N\right>\right).
\end{align}
\paragraph{Calculation of third order moment}
The third order moment of $\Delta V$ is given by
\begin{align}
\left<\Delta V^3\right> = \sum_{n=1}^{\infty} \frac{n^3+3n^2+2n}{\left(k_m/\alpha\right)^3}f_N(n)=\frac{\alpha^3}{k_{m}^3}\left(\left<N^3\right>+3\left<N^2\right>+2\left<N\right>\right).
\end{align}

When the burst size is one with probability one, we have $N=X$ with probability one. The formulas of mean, $CV^2$ and skewness of $\Delta V$ simplify to
\begin{align}
\left<\Delta V\right>=\frac{\alpha X}{k_{m}}, CV^2_{\Delta V}=\frac{1}{X}, skew_{\Delta V}=\frac{2}{\sqrt{X}}.
\end{align}
When the burst size is geometric, we get the following expressions:
\begin{align}
&\left<\Delta V\right>=\frac{\alpha}{k_{m}}\left(\frac{X}{b}+1\right), \\
&CV^2_{\Delta V}= \frac{var(\Delta V)}{\left<\Delta V\right>^2}=\frac{b^2+2 b X+X}{(b+X)^2}, \\
& skew(\Delta V)=\frac{2 \left(b^3+3 b^2 X+3 b X+X\right)}{\left(b^2+2 b X+X\right)^{3/2}}.
\end{align}

We also note that that the skewness of $\Delta V$ is positive in both cases considered above which is consistent with previous results \cite{vk97}.

\subsection*{Scale invariance of the distribution of $\Delta V$}
It has been shown in \cite{tbs15} that the distributions of the added volume $\Delta V$ in different growth conditions collapse when rescaled by respective $\left<\Delta V\right>$. Mathematically, we want to show that the probability density function $f_{\Delta V}(v)$ has the following form \cite[Supplementary Information equation 36]{tbs15}:
\begin{equation}
f_{\Delta V}(v) = \frac{1}{\left<\Delta V\right>} g \left( \frac{v}{\left<\Delta V\right>}\right),
\end{equation}
where $g(.)$ is an arbitrary normalized function. For the distribution in \eqref{eqn:distDeltaV}, let us consider the following function for $g$
\begin{equation}
g(w) = \sum_{n=1}^{\infty} \frac{\left(w \left<N\right>\right)^{n-1}}{(n-1)!} \left<N\right> \exp\left(-w\left<N\right>\right) f_N(n),
\label{eqn:g}
\end{equation} 
where $\left<N\right>$ is the expected value of the minimum number of transcription events required for the protein to cross the threshold $X$ and is given by $\left<N\right>=\frac{X}{b}+1$. Also, it is related with  $\left<\Delta V\right>$ as described in \eqref{eqn:meanDeltaV}.

For the function $g$ given in \eqref{eqn:g}, we have
\begin{align}
\frac{1}{\left<\Delta V\right>} g \left( \frac{v}{\left<\Delta V\right>}\right) &= \frac{1}{\left<\Delta V\right>} \sum_{n=1}^{\infty} \frac{\left(\frac{v\left<N\right>}{\left<\Delta V\right>}\right)^{n-1}}{(n-1)!} \left<N\right> \exp\left(-\frac{v\left<N\right>}{\left<\Delta V\right>}\right) f_N(n) \\
&= \sum_{n=1}^{\infty} \frac{\left(\frac{k_{m} v}{\alpha}\right)^{n-1}}{(n-1)!} \frac{k_{m}}{\alpha} \exp\left(-\frac{k_{m} v}{\alpha}\right) f_N(n) \\
&= f_{\Delta V}(v).
\end{align}
This establishes the scale invariance of the distribution $f_{\Delta V}(v)$.

One consequence of the scale invariance property of $f_{\Delta V}(v)$ is that the normalized moments $\left<\Delta V^j\right>/\left<\Delta V\right>^j$ are independent of the growth conditions \cite[Supplementary Information]{tbs15}. This can be checked as follows.

The $j^{th}$ order conditional moment of $\Delta V$ (conditioned with respect to $N$) is given by $j^{th}$ order moment of an Erlang distribution. Thus
\begin{align}
\left<\Delta V^j|N=n\right>&=\left(\frac{\alpha}{k_{m}}\right)^j \left(n (n+1)(n+2)\cdots (n+j-1)\right) \\
\implies \left<\Delta V^j\right> &=\left(\frac{\alpha}{k_{m}}\right)^j \left<N (N+1)(N+2)\cdots (N+j-1)\right>.
\end{align}
Therefore using \eqref{eqn:meanDeltaV}, we have 
\begin{align}
\frac{\left<\Delta V^j\right>}{\left<\Delta V\right>^j}=\frac{\left<\left(N (N+1)(N+2)\cdots (N+j-1)\right)\right>}{\left<N\right>^{j}}
\end{align}
which is independent of the growth rate. 

This fact can be used to show that statistical measures such as noise ($CV^2$) and skewness are independent of the growth rate. Take $CV^2$ for instance. It is defined as
\begin{equation}
CV^2_{\Delta V}=\frac{\left<\Delta V^2\right>}{\left<\Delta V\right>^2}-1. 
\end{equation}
By the scale invariance property, ${\left<\Delta V^2\right>}/{\left<\Delta V\right>^2}$ is independent of the growth rate $\alpha$. Thus, the noise $CV^2$ is also independent of $\alpha$. Similarly,  skewness is given by
\begin{align}
skew_{\Delta V}&=\frac{\left<{\Delta V}^3\right>-3 
\left<\Delta V\right>var(\Delta V)-\left<\Delta V\right>^3}{\left(var(\Delta V)\right)^{3/2}} \\
&=\frac{\frac{\left<{\Delta V}^3\right>}{\left<\Delta V\right>^3}-3 \left(\frac{\left < \Delta V^2\right>}{\left<\Delta V\right>^2}-1\right)-1}{\left(\frac{\left < \Delta V^2\right>}{\left<\Delta V\right>^2}-1\right)^{3/2}},
\end{align}
which is again independent of $\alpha$ by the scale invariance property. 

\section{Distribution of FPT}


FPT distribution can be obtained from (Equation S1.10) by solving
\begin{align}
f_{FPT}\left(t\right) & =\int_{0}^{\infty}f_{FPT | V_0}\left(t\left|v\right.\right)f_{V_{0}}\left(v\right)dv,
\end{align}
where $f_{V_{0}}(v)$ is the probability distribution of cell volumes at birth. Ignoring the errors in partitioning of volume, it can be seen that $f_{V_{0}}\left(v\right)\thickapprox f_{\Delta V}\left(v\right)$ \cite{tbs15}. Using this relation, the FPT distribution is given by

\begin{alignat}{1}
 f_{FPT}(t) & =k_{m}e^{\alpha t}\sum_{n=1}^{\infty}\sum_{l=1}^{\infty}\frac{f_{N}(n)f_{N}(l)\left(\frac{k_{m}}{\alpha}\right)^{n+l-1}\left(e^{\alpha t}-1\right)^{n-1}}{(n-1)!(l-1)!}\int_{0}^{\infty}v^{n+l-1}\exp\left(-\frac{k_{m}v}{\alpha}e^{\alpha t}\right)dv\\
 & =\sum_{n=1}^{\infty}\sum_{l=1}^{\infty}\frac{f_{N}(n)f_{N}(l)(n+l-1)!}{(n-1)!(l-1)!}\alpha\frac{\left(e^{\alpha t}-1\right)^{n-1}}{\left(e^{\alpha t}\right)^{n+l-1}}\\
 & =\sum_{n=1}^{\infty}\sum_{l=1}^{\infty}\frac{f_{N}(n)f_{N}(l)(n+l-1)!}{(n-1)!(l-1)!}\alpha\sum_{i=0}^{n-1}\left(\begin{array}{c}
n-1\\
i
\end{array}\right)\left(-1\right)^{i}e^{-\alpha t(l+i)}\text{.}
\end{alignat}

FPT moments can be written as 

\begin{alignat}{1}
\left\langle FPT^{j}\right\rangle  & =\frac{j!}{\alpha^{j}}\sum_{n=1}^{\infty}\sum_{l=1}^{\infty}\frac{f_{N}(n)f_{N}(l)(n+l-1)!}{(n-1)!(l-1)!}\sum_{i=0}^{n-1}\left(\begin{array}{c}
n-1\\
i
\end{array}\right)\frac{\left(-1\right)^{i}}{\left(i+l\right)^{j+1}}\\
 & =\frac{K_{j}}{\alpha^{j}}\text{.}
\end{alignat}

The normalized moments are

\begin{alignat}{1}
\frac{\left\langle FPT^{j}\right\rangle }{\left\langle FPT\right\rangle ^{j}} & =\frac{K_{j}}{K_{1}}\text{.}
\end{alignat} This implies, as shown for $\Delta V$, that the noise, skewness and
higher order moments are indendent of growth rate. Furthermore, if
we use the function

\begin{alignat}{1}
g\left(w\right) & =\sum_{n=1}^{\infty}\sum_{l=1}^{\infty}\frac{f_{N}(n)f_{N}(l)(n+l-1)!}{(n-1)!(l-1)!}K_{1}\sum_{i=0}^{n-1}\left(\begin{array}{c}
n-1\\
i
\end{array}\right)\left(-1\right)^{i}e^{-K_{1}w(l+i)}\text{,}
\end{alignat} then FPT distribution can be written as

\begin{alignat}{1}
f_{FPT}\left(t\right) & =\frac{1}{\left\langle FPT\right\rangle }g\left(\frac{t}{\left\langle FPT\right\rangle }\right)\text{.}
\end{alignat} Thus, the FPT distribution is also scale invariant. In \cite{tbs15} constant  $K_1$ is equal to $\log{2}$. We found that $K_1\approx 0.7\approx \log{2}$ for different distributions of burst size  $B$ and several values of threshold $X$.

\section{Theoretical constraints on the molecular mechanism underlying adder principle}
\begin{figure}[h]
\centering
\includegraphics[width=0.35\textwidth]{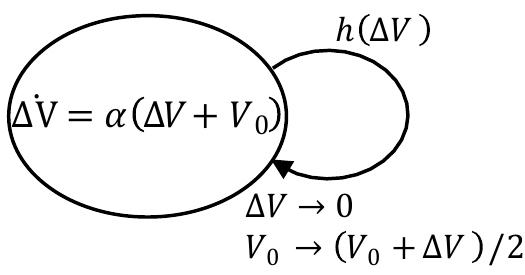}
\caption{Description of the cell division process as a stochastic hybrid system. The added volume $\Delta V$ evolves as per a deterministic dynamics until the division event takes place. The hazard rate for division is $h(\Delta V, \overline{\Delta V})$. Upon division, the added volume $\Delta V$ and the cell volume at birth $V_0$ reset to $0$ and $(V_0+\Delta V)/2$, respectively. }
\label{fig:shs}
\end{figure}

Let us consider a phenomenological description of the adder principle in terms of a hybrid system as shown in Fig.~\ref{fig:shs}. Assuming the volume of a cell at birth as $V_0$, the added volume $\Delta V$ follows a deterministic dynamics as $\dot{\Delta V}=\alpha\left(V_0+\Delta V\right)$. The hazard rate for division event is assumed to be some general function $h(\Delta V)$. The adder principle states that the division occurs, on average, when a constant volume has been added, irrespective of the initial cell volume $V_0$.

Using the Dynkin's formula \cite{dav93}, the expected value of the added volume $\left<\Delta V\right>$ follows
\begin{align}
\frac{d}{dt}\left\langle \Delta V\right\rangle = & \Big \langle \alpha\left(\Delta V+V_{0}\right)-h\left(\Delta V\right)\Delta V \Big \rangle\\
\approx & \alpha\left(\left\langle \Delta V\right\rangle +V_{0}\right)-h\left(\Delta V\right)\left\langle \Delta V\right\rangle.
\end{align}
We have made the mean-field approximation in the second equation above. In steady state, a solution which has $\left<\Delta V\right>$ independent of $V_0$ is only possible if the hazard rate $h\left(\Delta V\right)$ is a dirac delta function $h\left(\Delta V\right)=\delta \left(\Delta V - \overline{\Delta V}\right) $. Therefore any mechanism which actively senses the added volume has to trigger the division the moment it reaches the prescribed threshold. The time-keeper protein based mechanism proposed in this work satisfied this necessary condition, and since it produced the added model in distribution sense as well, it shows that this theoretical constraint is both necessary and sufficient. 

\section{Additional information on Fig. 2 in the main text}
To study the effect of volume at birth on the noise in generation time ($CV^2_{FPT}$) , we selected the rescaled mean and standard deviation of the generation time given the smallest and the largest volumes at birth from Fig. 2D in \cite{tbs15}. We assumed that the generation time has a normal distribution with mean and standard deviation given by the volume at birth (small and large). We drew 150 generation time samples for each volume at birth from the assumed distribution (150 is the number of samples per point used to generate plot in Fig. 2D). Using bootstrapping we tested if the $CV^2$ of the small cells is larger than the $CV^2$ of large cells. The p value obtained is $0.00023$. The 95\% confidence interval is given in the following table.
\begin{table}[h!]
\centering
\begin{tabular}{ c c c c}
\hline
{} & Lower bound & Median & Upper bound \\
Small volume & 0.0424 & 0.0545 & 0.0686 \\
Large volume & 0.0796 & 0.1030 & 0.1313 \\
\hline
\end{tabular}
\caption{\emph{$CV^2$ of timing increases with increase in cell size at birth}. Using the data from Fig. 2D in \cite{tbs15}, the bootstrapped values of $CV^2$ with $95\%$ CI are provided. }
\end{table}
\newpage
\begin{acknowledgments}
AS is supported by the National Science Foundation Grant DMS-1312926. The authors thank Prof. Suckjoon Jun for providing experimental data to compare with model predictions on noise in division time (Fig. 2).
\end{acknowledgments}
\bibliographystyle{unsrt}
\section*{References}
\bibliography{ref}
\end{document}